\documentclass{emulateapj}

\usepackage{amsmath,natbib,graphics,euscript} %,epsfig}

\slugcomment{To be Submitted to ApJ}

\begin{document}
%------------------------------------------------------------------------------

%\title{Effects of Radiative Cooling in General Relativistic Magnetohydrodynamic Simulations of Black-Hole Accretion Disks}
\title{General Relativistic Magnetohydrodynamic Simulations of the Hard State as a Magnetically-Dominated Accretion Flow}

\author{P. Chris Fragile}
\affil{Department of Physics \& Astronomy, College of Charleston,
Charleston, SC 29424; fragilep@cofc.edu}

\and

\author{David L. Meier}
\affil{Jet Propulsion Laboratory, California Institute of
Technology, Pasadena, CA 91109; David.L.Meier@jpl.nasa.gov}

\date{{\small    \today}}
\date{{\small   \LaTeX-ed \today}}
%-----------------------------------------------------------------------------

\begin{abstract}
We present one of the first physically-motivated two-dimensional general relativistic
magnetohydrodynamic (GRMHD) numerical simulations of a radiatively-cooled black-hole
accretion disk. The fiducial simulation combines a  total-energy-conserving formulation
with a radiative cooling function, which includes bremsstrahlung,
synchrotron, and Compton effects. By comparison with other simulations we show that 
in optically thin
advection-dominated accretion flows,  radiative cooling can
significantly affect the structure,
without necessarily leading to an optically thick, geometrically
thin accretion disk. We further compare the results of our radiatively-cooled simulation
to the predictions of a previously developed analytic model for
such flows. For the very low stress parameter and accretion rate
found in our simulated disk ($\alpha \approx 0.003$,
$\dot{M}/\dot{M}_{Edd} \approx 5 \times 10^{-6}$), we
closely match a state called the ``transition'' solution between an
outer advection-dominated accretion flow and what would be a
magnetically-dominated accretion flow (MDAF) in the interior. The
qualitative and quantitative agreement between the numerical and
analytic models is quite good, with only a few well-understood
exceptions. According to the analytic model then, at significantly
higher $\alpha$ or $\dot{M}$, we would expect a full MDAF to form.

The collection of simulations in this work also provide
important data for interpreting other numerical results in the literature, as they span the most common treatments of thermodynamics, including simulations evolving: 1) the internal energy only; 2) the internal energy plus an explicit cooling function; 3) the total energy without cooling; and 4) total energy including cooling. We find that the total energy formulation is a necessary
prerequisite for proper treatment of radiative cooling in MRI
accretion flows, as the internal energy formulation produces a large
unphysical numerical cooling of its own.  We also find that the
relativistic cooling functions must be handled carefully numerically
in order to avoid equally unphysical heating or cooling runaways.
\end{abstract}

\keywords{accretion, accretion disks --- black hole physics ---
galaxies: active --- MHD --- relativity --- X-rays: stars}

\section{Introduction}
\label{sec:intro}

The process by which turbulent accretion flows around black holes
develop large scale magnetic fields that can drive collimated jet
outflows is still poorly understood.  Much of what we know comes
from observations of X-ray binaries (XRBs) \citep[summarized
by][]{fbg04}, largely because the rapid variability in these
stellar-mass systems allow for observations of state changes on
timescales of days to at most a few years, whereas state changes are
rarely observed for AGN. In XRBs, jets are associated with both 
the Hard and Soft accretion states \citep{mcc06}, although their
characteristics in the two states are quite different. Hard state
jets are very steady, potentially lasting for weeks, whereas jets produced 
in the transition to the high accretion-rate Soft state can be 
explosive and short-lived.  One
possible interpretation of this phenomenology \citep{fbg04} is that
accreting black holes produce jets most of the time, with the jet
speed being a function of the accretion rate. Objects in the Hard
state produce slow ($v \sim 0.3 c$) jets, while Soft-state objects
produce fast jets with flat-space Lorentz factors $W\equiv
(1-v^2/c^2)^{-1/2}\sim 10$ that are comparable with jets in higher
luminosity AGN (e.g., FR II radio sources). The explosive jet, in
this picture, is simply the bow shock of a new fast jet interacting
with a previously-existing slow jet as the source's accretion rate
temporarily, and rapidly, increases toward the Eddington limit.

Recent numerical simulations of non-radiative magnetohydrodynamic
(MHD) flows that are unstable to the magneto-rotational instability
(MRI) have been shown to produce jetted outflows \citep{hkvh04,
mck06}. In those simulations the mechanism involves the development
of a magnetically-dominated region close to the black hole and
rotation axis, where the magnetic field %reconnects and 
orders itself into a helical, rotating structure that drives the jet. At the
present time, however, it is not clear if these results fit the
phenomenology described above. The problem is that non-radiative MRI
simulations should be the proper theoretical counterpart to the Hard
state, but the simulations produce relativistic jets instead of the
slow type jets associated with the Hard state.

One of
the present authors has suggested \citep{meier05, meier08, meier09}
that, at moderately low accretion rates (where turbulent,
advection-dominated accretion flows or ADAFs should occur), the
inflow inside a radius $R_1 \sim 100 \, r_G = 100 \, GM / c^2$
should develop a black hole magnetosphere structure similar to the
force-free ones studied recently by \citet{tt01, uzd04, uzd05}. In
this picture, closed field lines connecting the disk at $R_1$ with
the event horizon could funnel ionized plasma toward the black hole,
creating a magnetically-dominated accretion flow or MDAF. Open field
lines anchored near $R_1$, on the other hand, could drive a jet
outflow with a speed set by the dynamical timescale near $R_1$
[$\sim (GM / R_1)^{1/2} \sim 0.1 \, c$]. A further strength of this
model is that a relatively large magnetosphere might help explain
why quasi-periodic oscillations in the Hard state are observed in
the Hertz range, rather than kHz.

Our suggested mechanism for MDAF/magnetosphere formation is radiative
cooling in the previously-supposed, radiatively-inefficient ADAF.
Cooling would lower the plasma pressure and decrease the disk
vertical scale height, both of which lead to a dramatic increase
in the dominance of magnetic stresses (ratio of magnetic to gas
pressure greater than unity).  Strongly-magnetized plasmas are much
more stable to the MRI, leading to a decrease in turbulence and
a more ordered magnetic field.  This is precisely the same process
that occurred in the \citet{hkvh04} and \citet{mck06} simulations,
but now at a radius of $\sim 100 \, r_G$ instead of a few.

A critical assumption in this picture, though, is that the entire
plasma (electrons {\em and} ions) cools. This is in contrast to
current accretion theory, which asserts that whenever the flow
enters a hot, Hard X-ray state, the transfer of thermal energy
between ions and electrons is inefficient, leading to a
two-temperature, optically thin ADAF with cool electrons and hot
ions. However, some theoretical and numerical work has suggested
this may not be the case \citep[e.g.][]{beg88,sha07}. There is also
some observational evidence that efficient energy transfer from ions
to electrons must occur even when a black hole is in a Hard X-ray
state. Monitoring of many black-hole candidate sources shows that
they can be found in the Soft state at bolometric luminosities 
lower than in the maximum Hard state. This appears to be especially
true in Hard states where a strong, steady jet is produced (e.g.,
the plateau state). Sources at the top right of Fig. 7 of
\citet{fbg04} (the ``FBG diagram'') are quite hard and yet quite
luminous. As a concrete example, Cygnus X-1 produces 90\% or more as
much bolometric luminosity in the Hard state as in its Soft state
\citep{mcc02}. This is hardly ``radiatively inefficient'' accretion
by any standard. The only truly inefficient
state might be the Quiescent state, of which the black-hole
candidate A0620-00 and the Galactic center black hole Sgr A* may be
examples.

Our goal in this paper is not to resolve the controversy of whether
or not Hard state objects radiate efficiently, but rather to
investigate how they will behave if they do. We proceed by
performing general relativistic MHD simulations of MRI-unstable
black hole accretion flows, with two key differences from previous
investigations:  we include in the energy equation a plausible
high-temperature cooling function that is relevant for such flows,
and we assume that electrons and ions are sufficiently thermally
coupled that cooling of the former also cools the latter, keeping
$T_i \approx T_e$.  For completeness, we actually compare four
different classes of numerical models: one that evolves internal
energy without including cooling, one that evolves internal energy and includes cooling, one that conserves total energy
but does not include cooling, and one that conserves total energy
and includes cooling. These four classes of models span most of the
simulations that have been carried out to date by other authors, and expands significantly on what has been done thus far with simulations involving physical cooling mechanisms. Our
paper is unique in that it gives the first direct comparison of all
four using a single numerical code and which, we believe, is the
first to investigate numerically the triggering of %magnetic
state transitions in cooled black hole accretion flows.

\section{Numerical Methods}
\label{sec:methods}

This work is carried out using the Cosmos++ astrophysical MHD code
\citep{ann05}. Cosmos++ includes
several schemes for solving the MHD equations, including a
traditional artificial viscosity (AV) scheme and a new extended
artificial viscosity (eAV) method. The AV scheme is based on an
internal-energy-evolving (entropy-conserving) scheme, whereas the eAV scheme is a
hybrid dual energy scheme that solves both the internal and total
energy equations. The eAV scheme has the obvious advantage that it
conserves total energy; it is also potentially more accurate than
other fully conservative schemes in tracking the internal energy of
the gas because of the dual treatment. Artificial viscosity based
schemes, which both of these are, further have the advantage that
they are simpler to deal with when it comes to including extra
physics such as the radiative cooling being added in this work.
Furthermore, the combination of AV and eAV methods allows us to
directly compare, within a single numerical code, the effects of
including realistic heating and cooling processes in the evolution
of MRI turbulent accretion disks.

Cosmos++ has options to solve the MHD equations in either a
Newtonian or general relativistic framework. Here the general
relativistic form is used. In writing our equations we use the
standard notation in which four- and three-dimensional tensor
quantities are represented by Greek and Latin indices, respectively,
and repeated indices imply summation. The equations of mass
conservation, momentum conservation, and magnetic induction, common
to both numerical methods used in this work, have the form
\begin{eqnarray}
 \partial_t D + \partial_i (DV^i) &=& 0 ~,  \label{eqn:av_de} \\
 \partial_t S_j + \partial_i (S_j V^i) &=&
      \frac{1}{4\pi} \partial_t (\sqrt{-g} B_j B^0)
    + \frac{1}{4\pi} \partial_i (\sqrt{-g} B_j B^i) \\ \nonumber
    & & {} + \left( \frac{S^\mu S^\nu}{2S^0} - \frac{\sqrt{-g}}{8\pi}
             B^\mu B^\nu \right) \partial_j g_{\mu\nu} \\ \nonumber
    & & {} - \sqrt{-g}~\partial_j \left( P + P_B + Q \right) \\ \nonumber
 & & {} +\Gamma W u_j \Lambda ~, \label{eqn:av_mom} \\
 \partial_t \mathcal{B}^j + \partial_i (\mathcal{B}^j V^i) &=&
    \mathcal{B}^i \partial_i V^j + g^{ij}~\partial_i \psi ~,
      \label{eqn:av_ind} \\
 \partial_t \psi + c_h^2 \partial_i \mathcal{B}^i &=&
 -\frac{c_h^2}{c_p^2} \psi ~, \label{eqn:div_clean}
\end{eqnarray}
where $g_{\mu\nu}$ is the 4-metric, $g$ is the 4-metric determinant,
$W=\sqrt{-g} u^0$ is the relativistic boost factor, $D=W\rho$ is the
generalized fluid density, $V^i=u^i/u^0$ is the transport velocity,
$u^\mu = g^{\mu \nu} u_\nu$ is the fluid 4-velocity, $S_\mu = W(\rho
h + 2P_B) u_\mu$ is the covariant momentum density, $P$ is the fluid
pressure, $Q$ is the artificial viscosity used for shock capturing, $\Gamma$ (without subscripts or superscripts) is
the adiabatic index, 
$c_h$ and $c_p$ are coefficients to determine the strength of
the hyperbolic and parabolic pieces of the divergence cleanser,  and $\Lambda(\rho, T, H, B)$ is the cooling function of a
gas with density $\rho$, temperature $T$, temperature scale height
$H$, and magnetic field strength $B$, as described in detail in the
next section. (With indices, $\Gamma$ indicates the geometric
connection coefficients of the metric.)
There are two representations of the magnetic field in our
equations: $B^\mu$ is the 4-vector of the magnetic field, which can
be defined in terms of the dual of the Faraday tensor ($B^\mu \equiv
u_\nu {^*F^{\mu\nu}}$), and $\mathcal{B}^i = W(B^i - B^0 V^i)$ is the
boosted magnetic field 3-vector, where $B^0$ is recovered from the
orthogonality condition $B^\mu u_\mu = 0$,
\begin{equation}
B^0 = -\frac{W}{g} \left(g_{0i} \mathcal{B}^i + g_{ij} \mathcal{B}^j
        V^i \right ) ~. \label{eqn:B0}
\end{equation}
The magnetic pressure is $P_B = \vert\vert B \vert\vert^2/8\pi =
g_{\mu\nu}B^\mu B^\nu/8\pi$. We have assumed an equation of state of
the form $P=(\Gamma-1)\rho \epsilon$ with $\epsilon$ being the
internal energy. We use the scalar $Q$ from
\citet{ann05} with $k_q=2.0$ and $k_l=0.3$. We fix the divergence
cleanser coefficients to be $c_h = c_{\rm cfl} \Delta x_{\rm
min}/\Delta t$ and $c_p^2 = c_h$, where $c_{\rm cfl} = 0.5$ is the
Courant coefficient, $\Delta x_{\rm min}$ is the minimum covariant
zone length, and $\Delta t$ is the evolution timestep.

Both computational schemes also solve the internal energy equation
in the form
\begin{eqnarray}
\partial_t E + \partial_i (EV^i) & = &
    - P \partial_t W - \left(P + Q\right) \partial_i (WV^i) \\ \nonumber
    & & {} + W\Lambda(\rho, T, H, B) ~,
    \label{eqn:av_en}
\end{eqnarray}
where $E=We=W\rho\epsilon$ is the generalized internal energy
density. The temperature $T$ in the cooling function is recovered from the internal
energy of the gas using the ideal gas law
\begin{equation}
T = (\Gamma-1)e/(k n)
\end{equation}
where $n=\rho/(\mu m_H)$ is the number density of the gas and we use
$\mu=1.69$.

Additionally the hybrid dual energy scheme solves the following
total energy equation
\begin{eqnarray}
\partial_t {\cal E} & + & \partial_i \left({\cal E}V^i\right)
           = \Sigma^0 - \partial_i \left(F^i\right) \\ \nonumber
           & + & \left(\frac{W^2 \Gamma}{\sqrt{-g}} + \sqrt{-g}g^{00}(\Gamma-1)\right)\Lambda(\rho, T, H, B)~,
\label{eqn:dual_en}
\end{eqnarray}
where the total energy ${\cal E}$ is defined as
\begin{eqnarray}
{\cal E}   = \sqrt{-g} T^{00} & = & \frac{W^2}{\sqrt{-g}} (\rho h +
2P_B) + \sqrt{-g}~g^{00} (P+P_B) \\ \nonumber
& & {} - \frac{1}{4\pi} \sqrt{-g} B^0 B^0
~,
\end{eqnarray}
the curvature source term is
\begin{equation}
\Sigma^0 = -\sqrt{-g}T^{\alpha \beta} \Gamma^0_{\alpha \beta} ~,
\end{equation}
and the divergence flux contribution $F^i$ is defined as
\begin{eqnarray}
   F^i & = & \sqrt{-g}~\left ( (g^{0j} - g^{00} V^j)~((P+P_B) \delta^i_j + Q^i_j) \right. \\ \nonumber
       & & {} \left. - \frac{1}{4\pi}(B^i B^0 - B^0 B^0 V^i) \right)
       ~.
\end{eqnarray}
Ideally this total energy equation would be sufficient by itself.
However, total energy schemes can run into trouble when recovering
local values for the internal energy. Defining ${\cal E}_D$ as the
non-thermal or ``dynamical'' component of the conserved energy,
\begin{equation}
{\cal E}_D = \frac{DW}{\sqrt{-g}} + \frac{2 P_B W^2}{\sqrt{-g}}
           + \sqrt{-g}\left( g^{00} P_B - \frac{B^0 B^0}{4\pi}\right) ~,
\end{equation}
we write
\begin{equation}
\widetilde{E} = \frac{({\cal E}-{\cal E}_D)\sqrt{-g}~W}
                {\Gamma W^2 + (\Gamma-1)g^{00}(\sqrt{-g})^2}  ~,
\end{equation}
for the internal energy extracted from the conserved energy field.
The trouble arises when numerical truncation errors accumulate to
the point that the sum of different physical contributions exceed
the total energy (${\cal E}_D > {\cal E}$). This can occur in
kinematic or magnetic field dominated flows and in the vicinity of
strong shocks. The problem of negative energy can be avoided rather
simply by forcing a minimum threshold on $\widetilde{E}$ to
guarantee positivity. However, such a floor value is clearly not an
accurate representation of the internal energy. Here we can benefit
from having evolved the internal energy independently. We choose to
only use the internal energy extracted from the total energy field
whenever $\widetilde{E}>10^{-3} {\cal E}$. This avoids corrupting
the internal energy value with numerical truncation error.  The low
density, background gas can also create accuracy problems for the
total energy scheme because the density is occasionally reset to a
numerical floor value. Therefore, we further require ${\cal E} >
10^{-3} {\cal E}_\mathrm{max}$ as a condition for replacing $E$ with
$\widetilde{E}$. This effectively excludes the background gas.
Finally, to recover as much disk heating as possible we always use
the larger of $E$ or $\widetilde{E}$, provided the above two
conditions are met.

We find that the cooling timestep, $\Delta
t_\mathrm{cool}=c_\mathrm{cfl} e/\Lambda$, is generally much shorter
than the MHD timestep required for stability in the fluid evolution,
$\Delta t_\mathrm{MHD}=c_\mathrm{cfl} \Delta x/V$, where $\Delta x$
and $V$ are characteristic zone lengths and velocities,
respectively. Therefore, to save computational resources, we
subcycle the cooling calculation, updating the energy $E$, as well as the temperature $T$ and scale height $H$, in
each zone using only the cooling ``source'' term [final terms in
equation (\ref{eqn:av_en})] and a timestep
$\Delta t_\mathrm{cool}$ until a full MHD timestep is reached, i.e.
until $\sum_{i=1}^{N_\mathrm{steps}} (\Delta t_\mathrm{cool})_i =
\Delta t_\mathrm{MHD}$. Then we update the total energy $\cal{E}$ and momentum $S_j$ according to the final terms in equations  (\ref{eqn:dual_en}) and (\ref{eqn:av_mom}), respectively. After that we proceed with the next MHD update for
all other evolution terms using the normal timestep $\Delta
t_\mathrm{MHD}$. Occasionally we have to deal with cooling timesteps
that are unreasonably small due to very low temperatures or very
high cooling efficiencies, regimes well outside the normal limits.
To prevent the code from getting hung up at these points, we
restrict $N_\mathrm{steps}$ to be $\le100$. This limit is usually
applied only in regions of very low density or very low energy where
proper treatment of the fluid is inherently difficult.

\section{Cooling Function}
\label{sec:cooling}

Three cooling processes are treated in this work: bremsstrahlung,
synchrotron, and the inverse-Compton enhancement of each of these two.
Generally, we implement the equations of \citet{esi96},
with some changes.  Below we describe first the equations that we 
use when the radiation is optically thin and some modifications that 
are necessary in order for the cooling computations to work well in 
our numerical simulations.  We then describe the modifications 
necessary when the plasma becomes optically thick to these radiative 
processes.  In the extremely optically thick limit, the treatment 
is essentially the diffusion approximation.  

\subsection{Optically Thin Limit}

The total cooling rate for the optically thin gas is \citep{esi96}
\begin{equation}
q^- = \eta_{\rm br, C} \, q^-_{\rm br} + \eta_{\rm s, C} \, q^-_{\rm s} ~, \label{eqn:q-}
\end{equation}
where $q^-_{\rm br}$ and $q^-_{\rm s}$ are the bremsstrahlung and synchrotron 
cooling terms, respectively, and $\eta_{\rm br, C}$ and $\eta_{\rm s, C}$ are 
Compton enhancement factors.  The details of how 
to compute the Compton enhancements are given in \citet{esi96}.  Basically, 
$\eta(\nu)$ is a modified exponential function of the Compton parameter 
\begin{eqnarray*}
y & = & 4 ( \Upsilon + 4 \Upsilon^2) \, (\tau_{es} + \tau_{es}^2)
\end{eqnarray*}
where $\Upsilon \equiv k T_e/m_e c^2$ is the dimensionless electron temperature, 
and $\tau_{es}$ is the electron scattering optical depth.  $\eta(\nu)$ is limited to 
a maximum value of $3kT/h \nu$, where $h$ is Planck's constant. 
$\eta_{\rm br, C}$ is found implicitly by integrating $\eta(\nu) \, dq^-_{\rm br}/d\nu$ 
over appropriate frequencies, and 
$\eta_{\rm s, C}$ is approximated as $\eta(\nu_c)$, where $\nu_c$ is the 
critical frequency below which the synchrotron emission becomes self-absorbed. 
Note:  while we implement both enhancements, because 
synchrotron emission is dominant at temperatures where Comptonization becomes 
important, only $\eta_{\rm s, C}$ is important in our simulations.

The un-Comptonized bremsstrahlung cooling rate from \citet{esi96} is
\begin{equation}
q^-_{\rm br} = q_{ei}^- + q_{ee}^- + q_{\pm}^- ~,
\end{equation}
where
\begin{eqnarray}
q_{ei}^- & = & n_p (n_e + n_+) \times \\ \nonumber
& & {} \left\{
\begin{array}{lr}
    1.50 \times 10^{-22} \, \Upsilon^{0.5} \, (1+1.781\Upsilon^{1.34})  & \Upsilon<1 \\
    2.12 \times 10^{-22} \, \Upsilon \, [\ln (1.123 \Upsilon + 0.48) + 1.5]  & \Upsilon \ge 1
    \end{array} \right. \label{eqn:ei} \\
q_{ee}^- & = & (n_e^2 + n_+^2) \times \\ \nonumber
& & {} \left\{ \begin{array}{lc}
    2.56 \times 10^{-22} \, \Upsilon^{1.5} \, (1+1.1\Upsilon + \Upsilon^2  - 1.25\Upsilon^{2.5})  & \Upsilon < 1 \\
    3.42 \times 10^{-22} \, \Upsilon \, [\ln(1.123 \Upsilon) + 1.28]  & \Upsilon \ge 1
    \end{array} \right. \label{eqn:ee} \\
q_{\pm}^- & = & n_e n_+ \times \\ \nonumber
& & {} \left\{ \begin{array}{lc}
    3.43 \times 10^{-22} \, (\Upsilon^{0.5} + 1.7\Upsilon^2)  &  \Upsilon < 1 \\
    6.84 \times 10^{-22} \, \Upsilon \, [\ln(1.123\Upsilon) + 1.24]  & \Upsilon \ge 1
    \end{array} \right. \label{eqn:pm}
\end{eqnarray}
in units of $\mathrm{erg~cm}^{-3} \mathrm{~s}^{-1}$. These represent cooling due to electron-ion (\ref{eqn:ei}),
positron-ion (\ref{eqn:ei}), electron-electron (\ref{eqn:ee}),
positron-positron (\ref{eqn:ee}), and electron-positron
(\ref{eqn:pm}) processes. Here $n_p=n_e - n_+$ is the
number density of protons 
%, $n_e=n_p(1+n_+/n_p)$ is the number density of electrons, 
and $n_+$ is the number density of
positron-electron pairs. One can determine the ratio
$n_+/n_p=(n_+/n_e)/(1-n_+/n_e)$ needed to calculate some of these
terms using the following expression
\begin{eqnarray}
\frac{n_+}{n_e} & = & \frac{1}{\pi}
\left\{1+\left[ \frac{2\Upsilon^2}{\ln(1.12\Upsilon +1.3)} \right ] \right\} \, \times \\ \nonumber
 & & {} \left\{ \begin{array}{lc}
    2 \times 10^{-4} \Upsilon^{3/2} \exp (-2/\Upsilon)(1+0.015\Upsilon) & \Upsilon \ll 1 \\
    (112/27\pi)\alpha_f^2 (\ln \Upsilon)^3 (1 + 0.058/\Upsilon)^{-1} & \Upsilon \gg 1 ~,
    \end{array} \right. \label{eqn:n+}
\end{eqnarray}
where $\alpha_f$ is the fine structure constant.

The un-Comptonized synchrotron rate is a sum of optically thick and thin emission
\begin{equation}
q_s^- = \frac{2\pi k T}{H c^2} \int_0^{\nu_c} \nu^2 \, d\nu + \int_{\nu_c}^\infty \epsilon_s(\nu) \, d\nu ~,
\label{eqn:uncompt}
\end{equation}
where $H$ is the temperature scale height.  The critical frequency can be found by 
equating the optically thin and thick volume emissivities at $\nu_c$  
\begin{equation}
\epsilon_s(\nu_c) = \frac{2\pi}{H} \frac{\nu_c^2}{c^2} kT ~,
\label{eqn:nuc}
\end{equation}
and solving the above expression numerically. 
For an isotropic, full Maxwellian distribution of electrons and positrons, 
the optically thin volume emissivity is \citep{mny96, esi96}
\begin{eqnarray}
\epsilon_s(\nu, \vartheta) & = & 4.43 \times 10^{-30} \, 4 \pi \nu \, (n_e+n_+)
\, \times \\ \nonumber 
& & {} \frac{I\left(x_M/\sin \vartheta\right)}{K_2(1/\Upsilon)} 
\mathrm{~ergs~cm}^{-3} \mathrm{~s}^{-1} 
\label{eqn:eps_snu_approx}
\end{eqnarray}
where $x_M = \nu/\nu_M$ is the normalized frequency (with $\nu_M = 6.27 \times 
10^{18} \, B \, (kT)^2$ [cgs] being the critical {\em electron} frequency for 
a given temperature), $\vartheta$ is the angle between the observer and the 
magnetic field direction, and $K_{2}$ is the modified Bessel function of the 
second kind of order 2, given by the integral
\begin{equation}
K_2(1/\Upsilon) \equiv \frac{\Upsilon^2}{3} \, \int_{1/\Upsilon}^{\infty} (z^2 - 1/\Upsilon^2)^{3/2} \, e^{-z} \, dz
\label{eqn:k2}
\end{equation}
In the high-temperature limit, the electron-energy-integrated, unitless spectrum is 
given by the well-known expression
\citep{pac70} 
\begin{equation}
I\left(\frac{x_M}{\sin \vartheta}\right) \equiv \frac{\sin \vartheta}{x_M} \int_0^{\infty} z^2 \, e^{-z} \, F(x_M/z^2 \sin \vartheta) \, dz
\label{eqn:iofxm}
\end{equation}
where $F(x)$ is the normalized synchrotron spectrum for a single electron
\begin{eqnarray*}
F(x) & = & x \int_x^{\infty} K_{5/3}(\xi) \, d\xi
\end{eqnarray*}
\citet{esi96} further average equation (\ref{eqn:eps_snu_approx}) over $\vartheta$ to obtain 
the total emissivity needed in equations (\ref{eqn:uncompt}) and (\ref{eqn:nuc})
\begin{eqnarray}
\epsilon_s(\nu) & = & 4.43 \times 10^{-30} \, 4\pi \nu \, (n_e + n_+) \, \times \\ \nonumber
& & {} \frac{I' ( x_M )}{K_2(1/\Upsilon)} 
\mathrm{~ergs~cm}^{-3} \mathrm{~s}^{-1} ~, 
\label{eqn:epsilon}
\end{eqnarray}
The angle- and energy-integrated, unitless spectrum $I'(x_M)$ can be fit to the following expression
\citep{mny96}
\begin{eqnarray}
I' (x_M) & = & \frac{4.0505}{x_M^{1/6}} \, \left(1 \, + \, \frac{0.40}{x_M^{1/4}} \, + \, 
\frac{0.5316}{x_M^{1/2}} \right) \, \times \\ \nonumber
& & {} \exp(-1.8899 \, x_M^{1/3})
\label{eqn:iprimeofxm}
\end{eqnarray}
with no more than 2.7\% error over the range $0 < x_M < \infty$.

\subsection{Problems with the Cooling Functions}

Generally, equations (\ref{eqn:q-} -- \ref{eqn:nuc}) and (\ref{eqn:epsilon} -- \ref{eqn:iprimeofxm}) 
work fairly well in the temperature range $10^{8-11}$ K, which is the range over
which they were used by \citet{esi96}. However, this is not
sufficiently broad for our numerical simulations where the plasma
temporarily can attain very high or very low temperatures in a given
cell. In applying these equations over the temperature range
experienced in our simulations, we discovered the following
problems:
\begin{enumerate}
\item{The number of positrons and electrons (determined from equation \ref{eqn:n+})
diverges for $T > 2.4 \times 10^{11}$ K, causing the simulation to crash.}
\item{There is an error in the synchrotron cooling expression (equation 
\ref{eqn:epsilon}) that causes
unphysical enhancement of the emission below $T < 10^{8}$ K.
The error is so severe, that without a fix the simulations develop a
cooling runaway, which freezes the plasma into a cold, toroidally-dominated
magnetic state.}
\end{enumerate}

Our current fix for the first problem is very simple:  we ignore
positron cooling entirely (i.e., $n_+ = 0$). 
A complete fix to the positron/electron ratio calculation is being
investigated at present.  However, our cooled simulation generally
remains below $10^{11}$ K in most places, so our neglecting the
contribution of positron cooling is reasonably valid.

The second problem requires a more sophisticated solution.  The error in the
synchrotron cooling is caused by the use of {\em different lower 
integration limits} in equations (\ref{eqn:k2}) and (\ref{eqn:iofxm}). 
$K_2(1/\Upsilon)$ is the correct factor only if lower temperatures 
($\Upsilon \rightarrow 0$) are allowed in equation (\ref{eqn:iofxm}) 
also.  The result of this limit mismatch is that the 
denominator in equation (\ref{eqn:epsilon}) vanishes faster for 
low temperatures ($T < 10^8$ K) than the numerator, leading to
enormous cooling rates for plausible temperatures (in the range
$10^{5-7}$ K).  Such a problem would not have affected
\citet{esi96}'s results, which maintained temperatures above this
range. However, in a numerical simulation with many millions of
cooling computations over millions of cells and time steps, the
probability is quite high that such cool temperatures will be
attained somewhere in the flow, whereupon the entire structure will
catastrophically freeze.

Since we still wish to use equation (\ref{eqn:iprimeofxm}) for $I'(x_M)$, 
the best fix for this problem is simply to assume the same high temperature 
limit in equation (\ref{eqn:k2}) as was assumed in equation 
(\ref{eqn:iofxm}) (i.e., allow $1/\Upsilon \rightarrow 0$, which 
is equivalent to replacing $K_2(1/\Upsilon) \rightarrow 2\, \Upsilon^2$), 
resulting in a corrected synchrotron emissivity expression 
\begin{eqnarray}
\epsilon_s(\nu) & = & 4.43 \times 10^{-30} \, 2\pi \nu \, (n_e + n_+) \, \times \\ \nonumber
& & {} \frac{I' ( x_M )}{\Upsilon^2}
\mathrm{~ergs~cm}^{-3} \mathrm{~s}^{-1} ~, 
\label{eqn:epsilon_corr}
\end{eqnarray}
to be used instead of equation (\ref{eqn:epsilon}) in equation (\ref{eqn:uncompt}). 
The error introduced by using the high-temperature limit of $K_2$ is of the same order
as that caused by using a zero lower limit in the numerator of equation 
(\ref{eqn:iofxm}) (i.e., $I(x_M/\sin \vartheta)$) in the first place.  
And that error is negligible compared to the
total plasma emission, because it occurs at low temperatures where
bremsstrahlung dominates (see Fig. \ref{fig:cooling_function}).

Finally, a Saha ionization equation
is used to determine the electron density, which provides an exponential
cutoff in the cooling below $T \sim 10^4$ K.  Thus, we consider
continuum cooling only; no line or molecular cooling is included.

One cautionary comment about these cooling functions is worth noting.
The expressions contain many exponentials whose arguments easily can trigger
underflow or overflow in a digital computer.  If great care is not taken in
respecting these limitations, even correct coding of the cooling
functions will lead to non-physical results (heating or cooling runaways)
in the accretion flow.

\subsection{Optically Thick Limit}

To represent the cooling behavior in the optically thick limit, we use
a slightly modified version of \citet{esi96}'s equation (21) that is
suitable for multi-dimensional MHD simulations.  Our total cooling function 
is given by \citet{hub90} 
\begin{eqnarray}
\Lambda &=& \frac{-q^-}{1+\sqrt{3} \tau_{\rm abs} + \frac{3}{2} \tau \tau_{\rm abs}} 
\\
 & = & \frac{4\sigma T^4/H}{\frac{3}{2} \tau + \sqrt{3} + 1/\tau_{\rm abs}} ~,
\label{eqn:lambda_thick}
\end{eqnarray}
where the local temperature scale height is computed from 
\begin{equation}
H = \frac{T^4}{\vert \nabla \left(T^4\right) \vert} ~.
\label{eqn:H}
\end{equation}
This is a suitable definition for scale height in a multi-dimensional numerical 
simulation.  If the flow were to assume a thin disk structure, for example, equation (\ref{eqn:H}) 
would give the standard exponential (radiation energy density) scale height. 

The optical depth due to absorption is calculated as
\begin{equation}
\tau_{\rm abs} = \kappa_{\rm abs} \rho H
\end{equation}
with
\begin{equation}
\kappa_{\rm abs} = \frac{q^-}{4\sigma T^4 \rho} ~.
\end{equation}
And the total optical depth is computed using an averaged opacity
\begin{equation}
\tau = \left< \kappa \right> \rho H
\end{equation}
with the diffusion average of $\kappa$ being
\begin{eqnarray}
\left< \kappa \right> & 
\equiv & \frac{- 2 \, \vert \nabla (T^4) \vert^2}{\rho T^4
\nabla \cdot \left[ \nabla (T^4)/(\kappa \rho)  \right] } 
\\
& = & \frac{- 2 \, T^4}{\rho H^2 \nabla \cdot \left[ \nabla (T^4)/(\kappa \rho) \right]} ~.
\label{eqn:kappa_avg}
\end{eqnarray}
The total opacity used in this equation is given by $\kappa =
\kappa_{\rm abs} + \kappa_{\rm es}$ with $\kappa_{\rm es} = 0.4$
cm$^2$ g$^{-1}$ being the electron scattering opacity.
Note that our definitions for $H$ and $\left< \kappa \right>$ allow for redistribution of 
heat within an optically thick region ($\tau >> 1$);  equation (\ref{eqn:lambda_thick})
then reduces exactly to the diffusion approximation. 
Our approach also allows for non-local heating outside a marginally thick-thin 
transition region ($\tau \approx 1$) by photons from more optically thick regions, 
since the partially diffusive nature of equation (\ref{eqn:kappa_avg}) in this 
situation allows partial transport of heat from warmer to cooler regions.  However, 
in the optically {\em thin} case ($\tau << 1$ and $\Lambda = q^{-}$), there is 
{\em no} transport of heat from one region of the simulation to the other.  There 
is only heat loss from the plasma and, therefore, from the simulation grid.

\begin{figure}
\begin{center}
\includegraphics[width=0.45\textwidth]{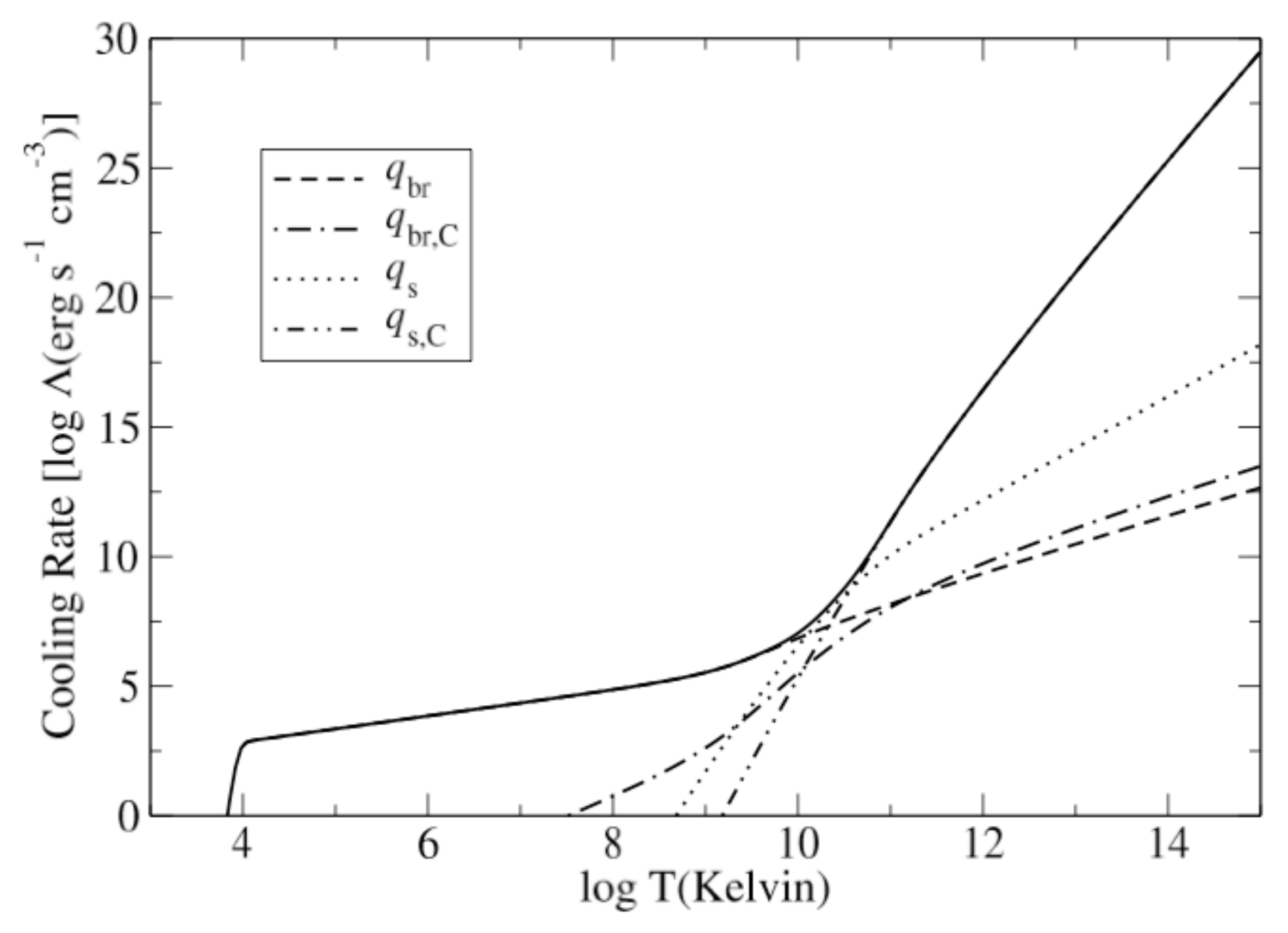}
\end{center}
%\plotone{f1.eps}
\caption{Plot of a sample cooling function with $\rho = 10^{-10}$ g
cm$^{-3}$, $B=8380$ G, and $H = 2.7 \times 10^7$ cm. The total cooling function
({\em solid} line) and the following components are represented:
bremsstrahlung ({\em short-dashed}); Compton enhancement to bremsstrahlung ({\em dot-long dash}); synchrotron ({\em dotted});
and the Compton enhancement to synchrotron ({\em dot-dot-dash}). 
\label{fig:cooling_function}}
\end{figure}

For the purpose  of illustration, in Figure
\ref{fig:cooling_function}, we plot the total cooling function
$\Lambda$ as a function of temperature, assuming fixed values of
$\rho$, $H$, and $B$. Of particular note is the temperature range
over which each of the cooling processes is important. For $10^4
\lesssim T\lesssim 10^9$ K, the dominant process is bremsstrahlung.
At higher temperatures bremsstrahlung no longer
dominates, even though its dependence with temperature steepens to
be proportional to $T \log T$. Instead, above $10^9$ K the dominant
cooling process is synchrotron, with Compton enhancement of
synchrotron becoming important for temperatures a little above that. Comptonization of bremsstrahlung, while included, is never particularly important
in our simulations.

%\clearpage

\section{Initialization}
\label{sec:initialization}

We initialize these simulations starting from the analytic solution
for a constant specific angular momentum torus around a non-rotating
black hole \citep{koz78}. In our initialization, the torus is
defined by its inner radius $r_\mathrm{in}=150 r_G$ and the radius of the
pressure maximum $r_{\rm center}=200 r_G$. Knowledge of $r_{\rm
center}$ leads directly to a determination of $\ell$, the specific
angular momentum of the torus, by setting it equal to the geodesic
value at that radius. Having chosen $r_\mathrm{in}$ we can obtain
$u_{in}=u_t (r_\mathrm{in})$, the surface binding energy of the torus, from
$u_t^{-2} = g^{tt}+\ell^2 g^{\phi\phi}$.

The solution of the torus variables can now be specified. The
internal energy of the torus is \citep{haw84a}
\begin{equation}
\epsilon(r,\theta) = \frac{1}{\Gamma} \left[
\frac{u_{in}}{u_t(r,\theta)} \right] ~.
\end{equation}
Thus the initial temperature of the torus, $T_0=(\Gamma-1)(\mu
m_H/k)\epsilon$, is fixed to be $\approx 10^9$ K by the
specification of the torus. Assuming an isentropic equation of state
$P=\rho \epsilon(\Gamma-1)=\kappa \rho^\Gamma$ for the
initialization, the density is given by $\rho = \left[
\epsilon(\Gamma-1)/\kappa \right]^{1/(\Gamma-1)}$. We take
$\Gamma=5/3$ and $\kappa=5 \times 10^{22}$ (cgs units). This gives
an initial density maximum in the torus of $\rho_{\rm
max,0}=2.8\times10^{-9}$ g cm$^{-3}$. Finally, the angular velocity
of the fluid is specified by
\begin{equation}
\Omega = V^\phi = -\ell \frac{g_{tt}}{g_{\phi \phi}} ~.
\end{equation}

Once the torus is constructed, it is seeded with a weak dipole
magnetic field in the form of poloidal loops along the isobaric
contours within the torus. The initial magnetic field vector
potential is \citep{dev03a}
\begin{equation}
A_\phi = \left\{ \begin{array}{ccc}
          b(\rho-\rho_{\rm cut}) & \mathrm{for} & \rho\ge\rho_{\rm cut}~, \\
          0                  & \mathrm{for} & \rho<\rho_{\rm cut}~.
         \end{array} \right.
\label{eqn:torusb}
\end{equation}
The non-zero spatial magnetic field components are then
$\mathcal{B}^r = - \partial_\theta A_\phi$ and
$\mathcal{B}^\theta =
\partial_r A_\phi$. The parameter $\rho_{\rm cut}=0.5*\rho_{\rm max,0}$ is used to keep the field a suitable
distance inside the surface of the torus. Using the constant $b$ in
equation (\ref{eqn:torusb}), the field is normalized such that
initially $\beta =P/P_B \ge \beta_0=10$
throughout the torus. The choice of the initial field geometry has
been shown to have relatively little effect on the development of
the MRI and the evolution of the disk \citep{bec08}, which is all we
are focused on in this manuscript. However, the initial field
topology does imprint itself in the formation and evolution of jets,
meaning that we will need to perform a more widely varying set of
simulations before addressing that topic.

In the background region not specified by the torus solution, we set
up a static, low density ($\rho = 10^{-6} \rho_{\rm max,0}$),
non-magnetic, hot gas. Numerical floors are placed on $\rho$ and $e$
at approximately $10^{-12}$ and $10^{-10}$ of their initial maxima,
respectively. The density floor is very seldom applied once the
initial background is replaced by evolved disk material. The energy
floor is applied somewhat more frequently. Nevertheless, these very
low floor values should not have any significant dynamical impact on
the problem.

These simulations are performed in 2.5 spatial dimensions (all three
spatial components of vector quantities are evolved, although
symmetry is assumed in the azimuthal direction) using a spherical
polar coordinate grid. The grid used in the majority of the
simulations consists of 192 radial zones and 128 zones in $\theta$.
We also performed select simulations at one-half and at double this
resolution to test the numerical convergence of our results. We find very little variation between our default resolution and the higher resolution simulation, suggesting our results are well converged. 

In the radial direction we use a logarithmic
coordinate of the form $\eta \equiv 1.0 + \ln (r/r_{\rm BH})$. The
spatial resolution near the black hole horizon is $\Delta r \approx
0.05 r_G$; near the initial pressure maximum of the torus, the
resolution is $\Delta r \approx 5 r_G$. Both are considerably
smaller than the initial characteristic MRI wavelength
$\lambda_\mathrm{MRI} \equiv 2\pi v_\mathrm{A}/\Omega \approx 50
r_G$. In the angular direction, we use a concentrated latitude
coordinate $x_2$ of the form $\theta = x_2 + \frac{1}{2} (1 - h)
\sin (2 x_2)$ with $h = 0.5$, which concentrates resolution toward
the midplane of the disk. As a result $r_{\rm center} \Delta \theta
= 4 r_G$ near the midplane while it is a factor of $\sim 3$ larger
for the zones near the pole.

For this work we have run the Cosmos++ numerical code in four
different modes: 1) internal-energy evolving with no explicit
cooling (model 522I or simply I); 2) internal-energy evolving including an explicit cooling function (model 522IC or simply I+C); 3) total-energy conserving with no
explicit cooling (model 522T or simply T); and 4) total-energy
conserving including an explicit cooling function (models 522TC or
simply T+C). The motivation for this is to allow for a clear, direct
comparison of simulations carried out under different physical
assumptions. The ``522'' in the long naming convention is a reference
to our choice of $\kappa=5 \times 10^{22}$.

\section{Results}
\label{sec:results}

Since no cooling processes are treated in simulations I and T, those
results simply scale with the mass of the black hole. However, for
purposes of comparison with simulations I+C and T+C, we will assume the same
scale for all variables in each simulation. Specifically we assume a
black hole mass of $M=10 M_\odot$, which sets the following physical
scales in the initial torus: $r_\mathrm{in} = 2.2\times10^8$ cm and $r_{\rm
center}=3.0\times10^8$ cm. The orbital period at $r=r_{\rm center}$
is $t_{\rm orb} = 1.77 \times 10^4 M = 0.875$ s. The initial gas
densities and temperatures are $\rho_{\mathrm{max},0} = 2.8\times
10^{-9}$ g cm$^{-3}$, $\rho_{\rm background} = 2.8\times 10^{-15}$ g
cm$^{-3}$, $T_{\rm disk} \approx 10^9$ K, and $T_{\rm background}
\approx 10^{11}$ K. The mass accretion rate is scaled by the
Eddington rate $\dot{M}_\mathrm{Edd} = 8.4\times 10^{18}$ g s$^{-1}$
for an $M=10 M_\odot$ black hole.  The models and parameters are summarized in Tables \ref{tab:models} and \ref{tab:params}. Each simulation is evolved for seven orbital periods. This is sufficient time for all four models to achieve approximate equilibriums inside $r\approx r_\mathrm{in}=150 r_G$. However, because these simulations are carried out in two dimensions, the anti-dynamo theorem prevents a true steady-state from being achieved, so these are only approximations of the true state.

\begin{deluxetable}{cc}
\tabletypesize{\scriptsize}
\tablecaption{Models \label{tab:models}}
\tablewidth{0pt}
\tablehead{
\colhead{Name} & \colhead{Description}
}
\startdata
I & Internal-energy evolving \\
I+C & internal energy + cooling \\
T & Total-energy conserving \\
T+C & Total energy + cooling \\
\enddata

\end{deluxetable}

\begin{deluxetable}{cc}
\tabletypesize{\scriptsize}
\tablecaption{Parameters \label{tab:params}}
\tablewidth{0pt}
\tablehead{
\colhead{Name} & \colhead{Initial value}
}
\startdata
$r_\mathrm{in}$ & $150r_G = 2.2\times10^8$ cm \\
$r_{\rm center}$ & $200r_G = 3.0\times10^8$ cm \\
$t_{\rm orb}$ & $1.77 \times 10^4 M = 0.875$ s \\
$\rho_{\mathrm{max},0}$ & $2.8\times 10^{-9}$ g cm$^{-3}$ \\
$\rho_{\rm background}$ & $2.8\times 10^{-15}$ g cm$^{-3}$ \\
$T_{\rm disk}$ & $\approx 10^9$ K \\
$T_{\rm background}$ & $\approx 10^{11}$ K \\
$M_{BH}$ & $10 M_\odot$ \\
$\dot{M}_\mathrm{Edd}$ & $8.4\times 10^{18}$ g s$^{-1}$ \\
\enddata

\end{deluxetable}

\subsection{Internal-Energy Evolving with No Cooling}

\begin{figure}
\begin{center}
\includegraphics[width=0.45\textwidth]{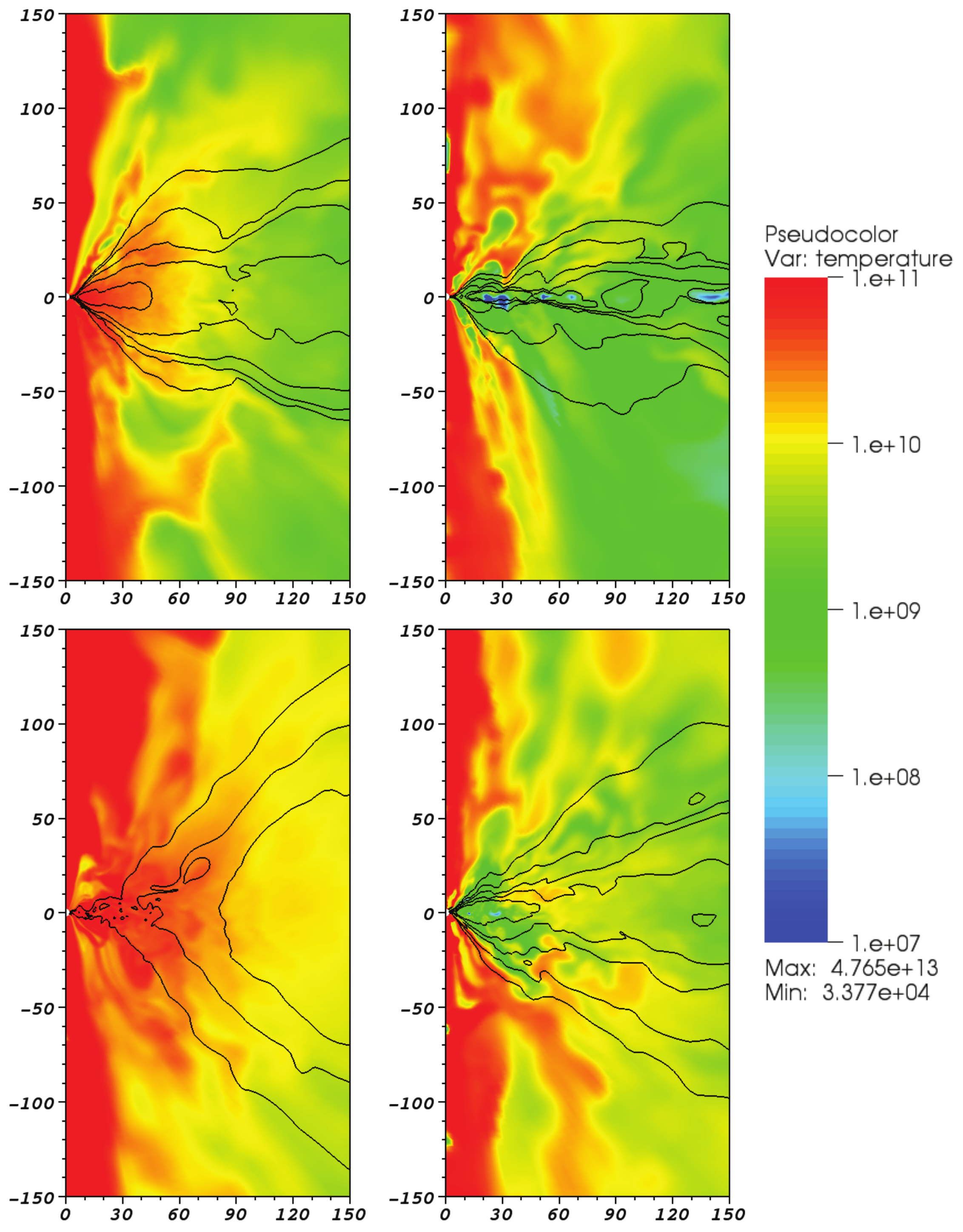}
\end{center}
%\plotone{f2.pdf}
\caption{Pseudo-color plots of $\log(T)$ with contours of
$\log{\rho}$. The {\em upper-left} panel 
is the final time dump of the internal energy model I; the {\em upper-right} panel
is the final time dump of the internal energy plus cooling model I+C; the {\em lower-left} panel is the final time dump of the total energy model T; and the {\em lower-right} panel is the final time dump of the total energy plus cooling model
T+C. The density contours are at $\rho = 0.005$, 0.016, 0.05, 0.16,
and $0.5 \rho_{\mathrm{max},0}$. \label{fig:T_rho}}
\end{figure}

In this simulation, which we designate 522I or just I, we only
evolve the internal energy equation (equation \ref{eqn:av_en}), ignoring the cooling function ($\Lambda=0$). This
mode of evolution has commonly been used in the past
\citep[e.g.,][]{dev03a,ann05}, particularly in codes derived from
the pioneering work of \citet{wil72}. From a thermodynamics
perspective, running the code in this mode is an interesting case
study. Because total energy is not conserved, any kinetic or
magnetic energy dissipated in the disk (except through shocks) is
simply lost from the simulation. In a sense, though, this creates a
sort of thermodynamic equilibrium, wherein cooling (in the sense of
energy lost from the disk) exactly matches dissipative heating
everywhere in the simulation. Thus, without explicitly treating
heating or cooling, this simulation actually mimics one that
includes heating plus a highly efficient cooling process. Some
caution is in order, though, in making such a statement. Some
heating and cooling mechanisms {\em are} captured in equation
(\ref{eqn:av_en}), specifically shock heating (through the artificial viscosity term) and adiabatic heating
and cooling. These may not be balanced in the same way they would in
a simulation that rigorously treated both heating and cooling.
Furthermore, this treatment implies rapid, efficient cooling
throughout the computational domain, {\em regardless of physical
conditions}. We will explore this point further in \S \ref{sec:tmc}.
In the {\em upper-left} panel of Figure \ref{fig:T_rho}, we plot the final distribution of
gas density and temperature for this model.

\subsection{Internal-Energy Evolving with Cooling}

For this simulation, which we designate 522IC or simply I+C, we again evolve the internal-energy equation (equation \ref{eqn:av_en}), this time including the radiative cooling term. Physically speaking, there is relatively little motivation for this model, as we know there are important dissipative heating processes in disks that are ignored in this model. Nevertheless, this model does serve to round out our small lattice of tests and demonstrate the importance of using a fully conservative energy scheme (or some other heat-capturing procedure) when including radiative cooling processes. This is because, without including heating, there is nothing to counterbalance the cooling, and the disk ends up unreasonably cold and thin, with temperatures below $10^7$ K over much of the disk midplane, as shown in the {\em upper-right} panel of Figure \ref{fig:T_rho}. 

\subsection{Total-Energy Conservation with No Cooling}

For this simulation, which we designate 522T or simply T, we use the
total energy conserving mode of Cosmos++ (again with $\Lambda=0$). By evolving equation
(\ref{eqn:dual_en}) and conserving total energy, we effectively
capture dissipative heating mechanisms ignored in the previous
simulations since any losses to the kinetic or magnetic energy of the
gas are recovered as heat. Total energy conserving codes \citep[as used
previously by e.g.,][]{mck06,noble_07}, are particularly
applicable when considering radiatively inefficient accretion flows
(RIAFs), such as the one that is thought to be currently feeding Sgr
A* \citep{nar95,yuan_03}. Because these disks (both simulated and
real) are not able to radiate their heat away efficiently, they tend
to be very hot and vertically thickened, as shown in the {\em lower-left} panel of Figure
\ref{fig:T_rho}.

\subsection{Total Energy Conservation with Cooling}
\label{sec:tmc}

For this simulation, which we designate 522TC or T+C, we use the
dual energy evolving mode of Cosmos++ described in \S
\ref{sec:methods}. By including equation (\ref{eqn:dual_en}) and
conserving total energy, we again effectively capture dissipative
heating mechanisms in the disk. In addition to the total energy
equation, we simultaneously evolve the internal energy equation
(\ref{eqn:av_en}) to ensure we recover reasonable values for the
internal energy whenever the total energy budget is dominated by
non-thermal components. This is particularly important for
calculating the temperature of the gas, which is a crucial input
into the cooling routine. This is the only simulation where the dissipative heating processes are balanced by a physically motivated local cooling function, as described in \S
\ref{sec:cooling}. As expected, this leads to an intermediate disk state between the hot, thickened RIAF state of simulation T and the unrealistically cooled disk in simulation I+C. The results are shown in the {\em lower-right} panel of Figure \ref{fig:T_rho}.

\section{Comparison of Numerical Models}

Simply looking at Figure \ref{fig:T_rho} and comparing the four models, we already note a number of qualitative differences. Obviously the disk in model T, which is expected to capture heating appropriately but includes only adiabatic cooling, is {\em much} hotter and thicker than any of the other three simulations. This is consistent with the expectations of a radiatively inefficient, two-temperature gas, where the ions are poorly coupled to the electrons. The opposite extreme is model I+C, which includes radiative cooling processes without capturing most of the real, physical heating in the disk. This leads to a very thin, cold disk solution, which could only apply in cases of very weak turbulence or very low ionization.

More interesting is to compare models I, the internal-energy evolving model, and T+C, the total energy plus cooling model. As we said before, model I can be thought of as a radiatively efficient model, but an unphysical one where cooling equals heating practically everywhere in the flow. This gives a much cooler disk than in model T, but also one in which the temperature increases monotonically as gas moves radially inward through the disk (compressive heating becomes more important). This is in contrast to model T+C, which shows an approximately constant or even slightly decreasing temperature for $r<150r_G$, due to the efficiency of Compton enhanced synchrotron radiation. Next we make a more quantitative comparison of the models.

\subsection{Angle-Averaged Properties of the Simulations}

\begin{figure}
\begin{center}
\includegraphics[width=0.45\textwidth]{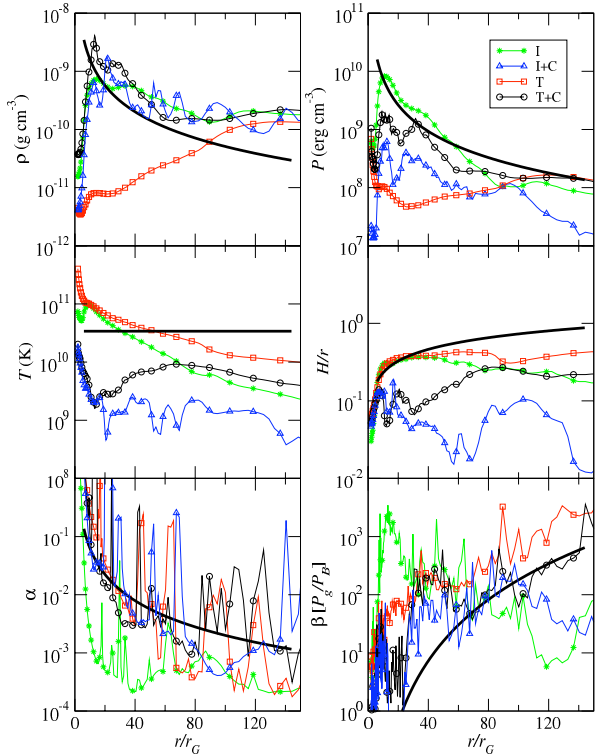}
\end{center}
%\plotone{f3a.eps}
\caption{Main disk properties plotted as a function of radius for
the internal-energy model I, the internal-energy plus cooling model I+C, the total-energy model T, and the total-energy plus cooling model T+C. The data have been time-averaged over
the final two orbital periods of each simulation. $P$, $T$, $\alpha$, and
$\beta$ are density-weighted averages. The thick solid line in each
frame is the solution for the MDAF transition region from equations
(\ref{eqn:mdaf}). \label{fig:disk_properties}}
\end{figure}

First, we construct
density-weighted spherical shell averages of the various disk
properties.  The formula we use is
\begin{equation}
\langle\mathcal{Q}\rangle_A(r,t) = \frac{1}{A} \int^{2\pi}_0
\int^\pi_0 \mathcal{Q} \sqrt{-g}
\, \mathrm{d}\theta \, \mathrm{d}\phi ~,
\end{equation}
where $A = \int^{2\pi}_0 \int^\pi_0 \sqrt{-g} \, \mathrm{d}\theta
\, \mathrm{d}\phi$ is the surface area of the shell.  We also
average these quantities over the final two orbital periods of the
simulations, $5t_{\rm orb} = t_{\rm min} \le t \le t_{\rm max} =
7t_{\rm orb}$, to negate any transient features. The time averages
are defined as
\begin{equation}
\langle\mathcal{Q}\rangle_t = \frac{1}{t_{\rm max} - t_{\rm min}}
\int^{t_{\rm max}}_{t_{\rm min}} \mathcal{Q} \, \mathrm{d}t ~.
\end{equation}
The numerical results for the internal-energy model I, the internal-energy plus cooling model I+C, the total-energy model T, and the total-energy plus cooling model T+C are
shown in Figure \ref{fig:disk_properties}. We have also included the
predictions for the transition state solution, which we discuss below (Section 
\ref{sec:trans}). 

%\clearpage

%\clearpage
\begin{figure}
\begin{center}
\includegraphics[width=0.45\textwidth]{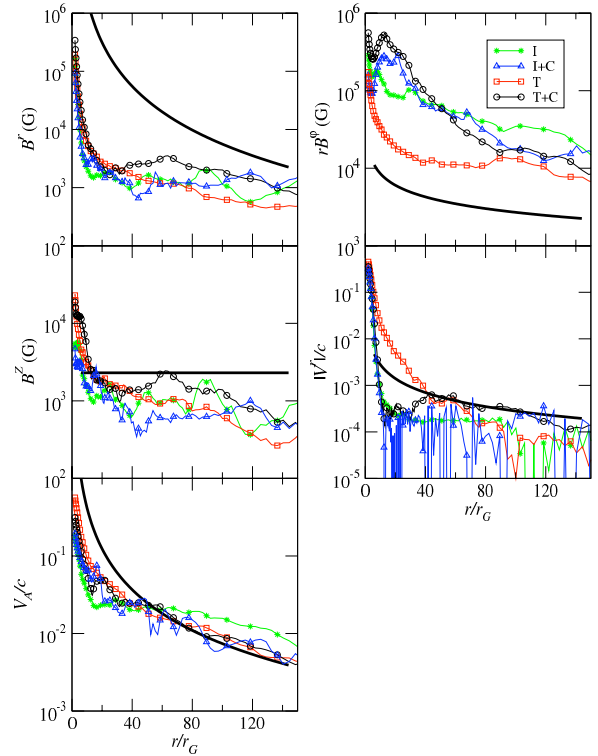}
\end{center}
%\plotone{f3b.eps}
\caption{Fig. \ref{fig:disk_properties} continued. $B^r$, $B^\phi$,
$B^Z$, $V^r$, and $V_A$ are density-weighted averages.
\label{fig:disk_properties_2}}
\end{figure}

\begin{figure}
\begin{center}
\includegraphics[width=0.45\textwidth]{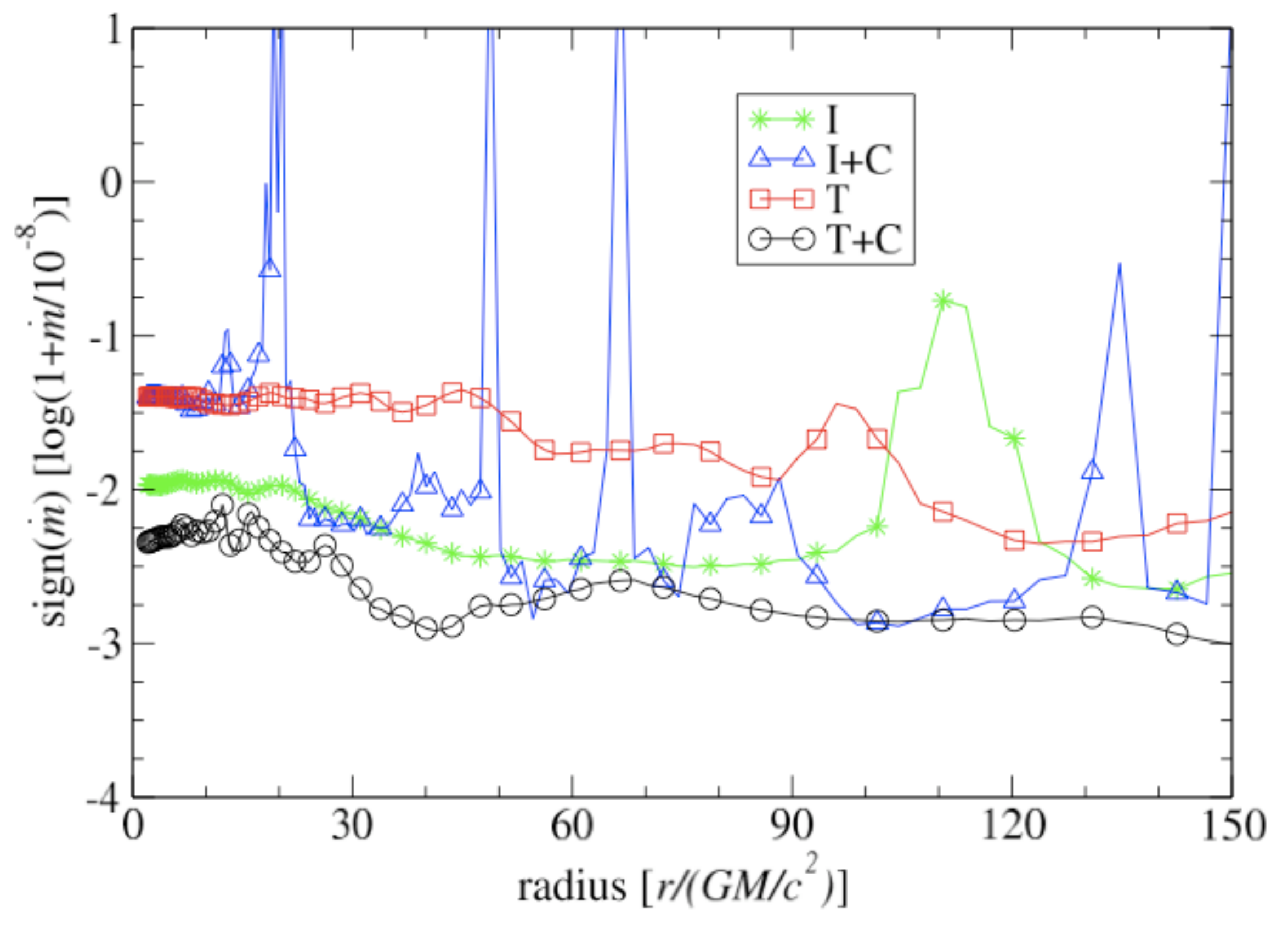}
\end{center}
%\plotone{f4.eps}
\caption{Plot of $\mathrm{sign}(\dot{m}) \log (1+\dot{m}/10^{-8})$ as a function of radius for the
internal-energy model I, the internal-energy plus cooling model I+C, the total-energy model T, and the total-energy plus cooling model T+C, where $\dot{m}=\dot{M}/\dot{M}_\mathrm{Edd}$. The data have been time-averaged over
the final two orbital periods of each simulation. By this time all three
simulations have achieved a reasonably steady inflow solution for $r\lesssim
150 r_G$. \label{fig:massflux}}
\end{figure}

There are clear outliers among the various disk models. For instance, the total-energy conserving model T exhibits significantly lower density, pressure, and azimuthal magnetic field in the inner regions than any of the other three simulations. This actually owes to its much lower accretion rate (shown in Figure \ref{fig:massflux}); material is just not moving through the disk very quickly. This, coupled with the considerably larger thickness of model T, leads to very low density and pressure. The internal-energy plus cooling model I+C is an outlier in the other direction, being an order of magnitude cooler and thinner than model T. Models I and T+C, on the other hand, representing unphysical and physical cooling, respectively, look very similar in many regards. In fact, the only notable exceptions are in $T$, $\alpha$, and $\beta$. We mentioned the difference in $T$ above, which is owing to the efficiency of Compton-enhanced synchrotron cooling for $T\gtrsim10^{10}$ K, and will return to the differences in $\alpha$ and $\beta$ below.

\subsection{Volume-Integrated Properties of the Simulations}

\begin{figure}
\begin{center}
\includegraphics[width=0.45\textwidth]{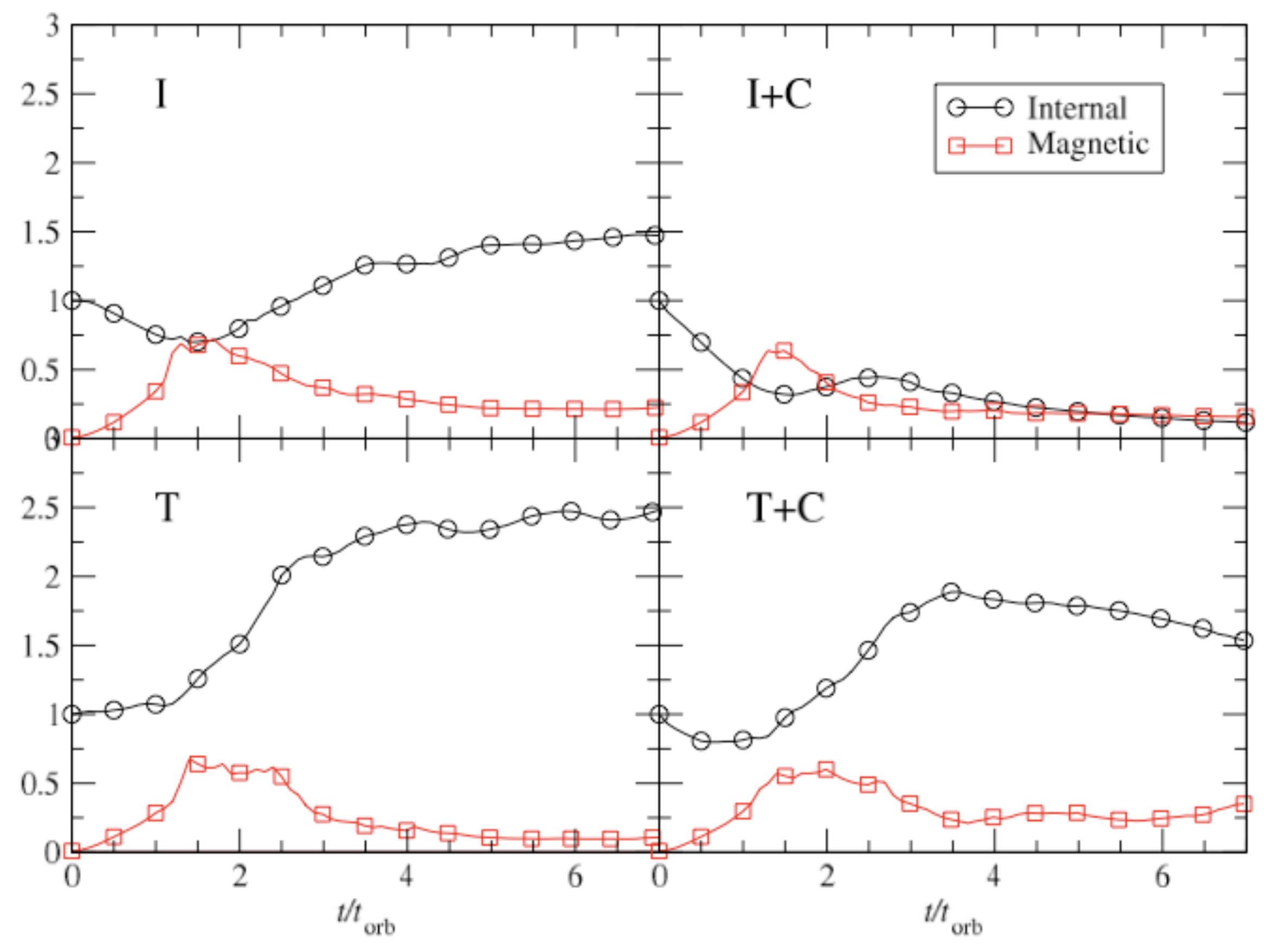}
\end{center}
%\plotone{f5.eps}
\caption{Plots of total internal ({\em circular symbols}) and
magnetic ({\em square symbols}) energies as functions of time for
the internal-energy model I ({\em top-left panel}), the internal-energy plus cooling model I+C ({\em top-right panel}), the total-energy
model T ({\em bottom-left panel}), and the total-energy plus cooling
model T+C ({\em bottom-right panel}). The energy scales have been
normalized to the initial internal energy in the
simulations.\label{fig:energy}}
\end{figure}

In Figure \ref{fig:energy} we plot the total integrated internal and
magnetic energies for each of our four classes of models. In the
magnetic energy we can see the characteristic growth of the
magneto-rotational instability on an orbital timescale, after which
it saturates. After about 2-3 orbital periods the magnetic energy
begins to decay due to accretion into the black hole, advection off
of the grid through the action of jets and winds, and also due to
the Cowling anti-dynamo theorem (the magnetic field is not able to
regenerate itself in two dimensions).

%\clearpage

Initially there is very little heating in the disk as the
magneto-rotational instability has not had time to build up the
turbulence and the flow is mostly laminar. Thus, in models I (which
fails to capture heating realistically), I+C (which neglects heating, yet includes cooling), and T+C (which captures heating, yet also includes cooling), the disk initially begins to cool. Once the MRI
really kicks in after about 1 orbit, models I, T, and T+C begin
heating, although model I heats more slowly than the two total
energy conserving models. Model I+C {\em never} shows significant heating, clearly demonstrating that the local cooling always dominates. Once heating begins, the non-radiative
models I and T never stop heating, whereas cooling
appears to catch up with heating in the radiatively-cooled total-energy model T+C
after about 4 orbits.

The differences between the internal energy curves of models I and T in Figure \ref{fig:energy} give some indication of the amount of energy simply lost from the simulation by model I, which uses only the internal energy formulation. This amount of energy is comparable to the amount of internal energy initially contained in the simulation. Likewise, the difference between the internal energy curves of models T and T+C say something about the level of cooling in the disk. The physical cooling in model T+C is of the same magnitude as the unphysical cooling of model I, but as we have already shown, the resulting disk structure has significant quantifiable differences.

We note here that the internal energy in the {\em lower-left} panel of Figure \ref{fig:energy} (model T) continues to increase all the way to the end of the simulation. This suggests that dissipative heating of the disk has not yet been fully quenched by anti-dynamo processes, so the cooling seen in model T+C is genuine.

\section{Comparison of Numerical and Analytic Results}
\label{sec:mdaf}

\begin{figure}
\begin{center}
\includegraphics[width=0.45\textwidth]{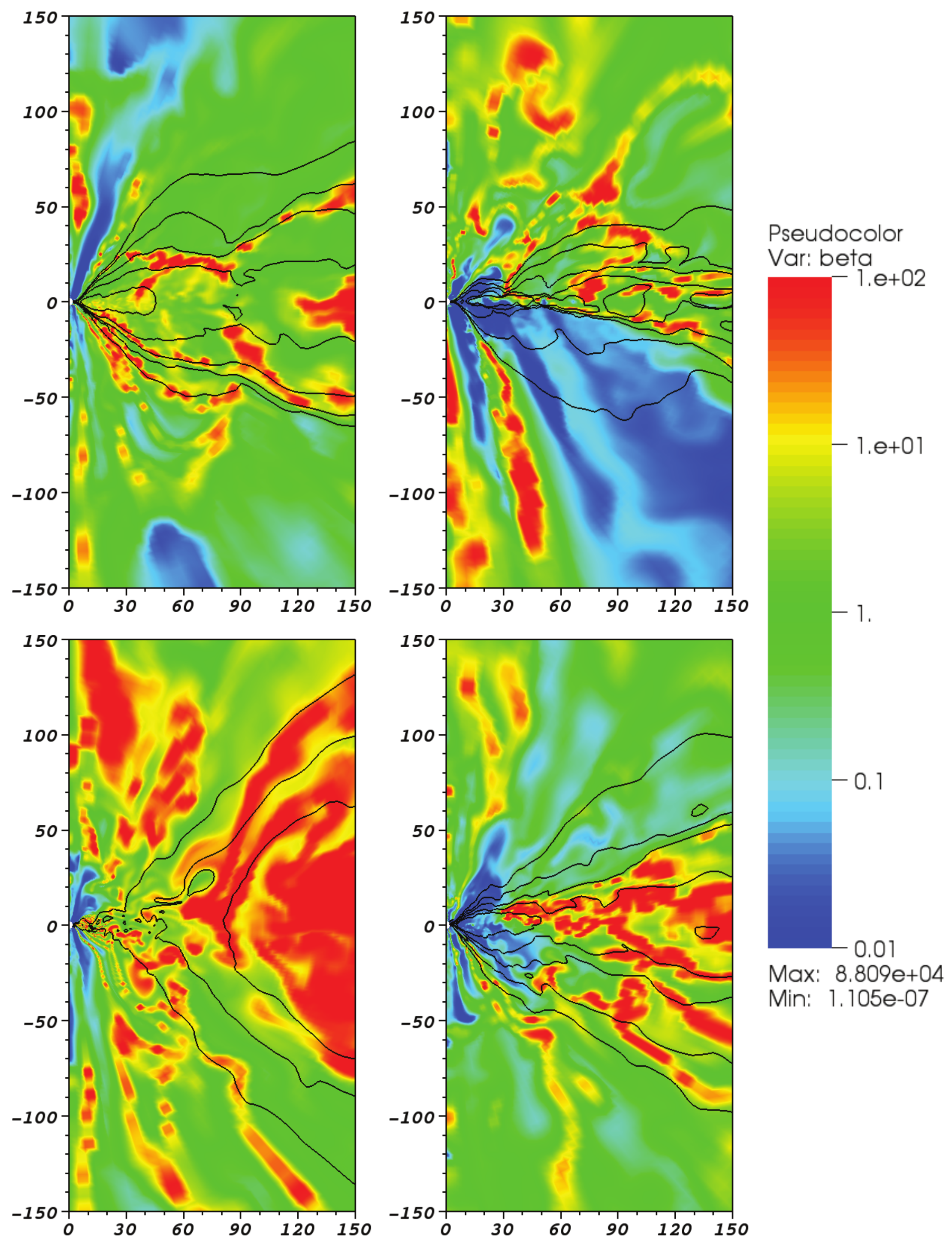}
\end{center}
%\plotone{f6.eps}
\caption{Pseudo-color plots of $\beta$ with contours of
$\log{\rho}$. {\em Panel a} 
is the final time dump of the internal-energy model I; {\em Panel b} is the final time dump of the internal-energy plus cooling model I+C; {\em Panel c}
is the final time dump of the total-energy model T; and {\em Panel
d} is the final time dump of the total-energy plus cooling model
T+C. The density contours are as in Fig. \ref{fig:T_rho}.
\label{fig:beta_rho}}
\end{figure}

Ultimately we would like to make direct comparisons between our 
numerical results and observations of black-hole accretion disks 
in the Hard state. In the meantime we can also compare our numerical 
results with applicable analytic work. An interesting comparison can 
be made between our radiatively-cooled model T+C and the MDAF model. The
basic idea of the MDAF is that catastrophic cooling in the inner
region of the disk should cause the disk to collapse vertically and
dramatically reduce the thermal energy relative to the magnetic,
such that $\beta = P_{\rm gas}/P_B$ becomes $< 1$
\citep{meier05, meier08}. It is clear from Figure \ref{fig:T_rho}
that the radiatively cooled disk (T+C) indeed is much thinner than
the uncooled disk (T). However, it is apparent from Figure
\ref{fig:beta_rho}, where we plot $\beta$ over much of the domain of the simulation that although $\beta$ is
significantly lower in some regions of the radiatively cooled disk
relative to the uncooled disk (particularly for $r<100r_G$), its angle average is not less than unity at any radius. 
% it is not less than unity except in a few isolated 
% locations inside the disk.

Apparently we did not achieve a fully magnetically dominated state. Nevertheless, we can
make a quantitative comparison between our numerical
model T+C and the predictions of the MDAF model. Prior to becoming
magnetically dominated, the model predicts that the inflow should
pass through a ``transitional'' state, in which $\beta$
decreases from its initially large value to of order unity. It is
this transitional inflow solution, then, that we wish to compare
with our numerical simulation.

\subsection{Analytic Theory of Transitional Flow}
\label{sec:trans}

Analytic development of MDAF theory begins with the development of
a simple analytic model for the ADAF structure in the region where
the ion and electron temperatures are definitely equal (i.e.,
$r \gtrsim 144 \, r_G$).  The simple ADAF model is constructed
in a manner similar to the \citet{sha73} $\alpha$-model, except that
cooling of the flow is performed by an advective term [$Q_{adv} \approx
\dot{M} P / (2 \pi \, r^2 \, \rho)$] instead of the usual radiative term. The
Compton parameter in this optically thin flow
\begin{eqnarray*}
y \; \approx \; 16 \, \Upsilon^2 \, \tau_{es}
\end{eqnarray*}
remains less than unity for $r > R_0$, where
\begin{equation}
R_0 = 2.75 \times 10^8 \alpha^{-2/5} m \dot{m}^{2/5} \mathrm{~cm.}
\end{equation}
is defined as the radius where $y = 1$ in the ADAF,
with $m=M/M_\odot$, and $\dot{m}=\dot{M}/\dot{M}_\mathrm{Edd}$.
Compton cooling is unimportant until the inflow approaches this radius.
Note that the electron scattering optical depth $\tau_{es} = \kappa_{es} \,\rho \,H$
uses the electron scattering opacity $\kappa_{es}$ and disk scale height $H$.
Because we use spherical geometry, the scale height is equivalent to
\begin{eqnarray*}
H & \equiv & r \, \sin \, \Theta
\end{eqnarray*}
where $\Theta$ is the disk {\em angular} scale height, with most of the accretion
flow occurring in a polar angle range of $\pi / 2 - \Theta < \theta < \pi/2 + \Theta$.

Inside $R_0$, $y$ must be $ \geq 1$, Compton cooling becomes important, and the
transitional flow begins.  In fact, the analytic MDAF models
assume that generic Compton cooling is dominant, so $y \approx 1$ must
be true for $r < R_0$ \citep{sle76}. This cooling decreases the plasma temperature
in the transition region, and therefore the disk scale height,
which also enhances  $B^r$ and $rB^{\phi}$ by compression.  Conservation
of magnetic flux in the steady MHD inflow requires that
\begin{eqnarray*}
B^r & \propto & r^{-1} \, H^{-1}
\\
rB^{\phi} & \propto & (V^{r})^{-1} \, H^{-1}
\\
B^{Z} & \propto & (V^{r})^{-1} \, r^{-1}
\end{eqnarray*}
This means that the magnetic viscosity stress parameter $\alpha$ no
longer can remain uniform with radius
\begin{eqnarray*}
\alpha(r) & = & \frac{t^{\phi r}}{P} \; \propto \frac{rB^{\phi} \, B^r}{P} \;
\propto \; \left( \frac{r}{H} \right)^3
\end{eqnarray*}
Since $H \propto r^{3/2}$ in this solution,
the magnetic stress relative to the pressure must {\em increase} as the
inflow approaches the black hole ($\alpha \propto r^{-3/2}$).
The end of transitional flow, and
beginning of true MDAF flow, begins inside the radius
\begin{equation}
R_1 = 2.75 \times 10^8 \alpha^{4/15} m \dot{m}^{2/5} \mathrm{~cm,}
\end{equation}
where $\alpha(r)$ becomes unity. For model T+C, we find $\alpha
\approx 0.003$, $m=10$, and $\dot{m} \approx 5 \times 10^{-6}$, so
$R_0 \approx 144 r_G$ and $R_1 \approx 3 r_G$. We therefore do not
actually expect a full MDAF solution for this particular model.
Instead we can make comparisons with the transition solution that
applies between $R_1$ and $R_0$.

The analytic transitional flow model
model predicts the following set of scaling relations for the
disk properties \citep{meier05, meier08}:
\begin{eqnarray}
\rho_c & = & 2.1 \times 10^{-5} ~\alpha^{-1}  ~m^{-1} ~\dot{m} ~x^{-3/2} \mathrm{~g~cm^{-3}} \nonumber \\
P_c     & = & 7.4 \times 10^{12} ~\alpha^{-3/5} ~m^{-1} ~\dot{m}^{3/5} ~x^{-3/2} \mathrm{~erg~cm^{-3}} \nonumber \\
T_c      & = & 2.65 \times 10^9     ~\alpha^{2/5}  ~\dot{m}^{-2/5} \mathrm{~K} \nonumber \\
H          & = & 4.4 \times 10^4    ~\alpha^{1/5} ~m ~\dot{m}^{-1/5} ~x^{3/2} \mathrm{~cm} \nonumber \\
\alpha(x) & = & 5.5 \times 10^3 ~\alpha^{2/5} ~\dot{m}^{3/5} ~x^{-3/2} \nonumber \\
\beta(x) & = & 3.8 \times 10^{-9} ~\alpha^{2/5} ~\dot{m}^{-7/5}
~x^{7/2} \nonumber \\
B^r & = & 2.21 \times 10^{11} ~\alpha^{-1/2} ~m^{-1/2} ~\dot{m} ~x^{-5/2} \mathrm{~G} \nonumber \\
rB^\phi & = & 2.3 \times 10^6 ~\alpha^{3/10} ~m^{-1/2} ~\dot{m}^{1/5} ~x^{-1/2} \mathrm{~G} \nonumber \\
B^Z & = & 1.31 \times 10^5 ~\alpha^{1/2} ~m^{-1/2} \mathrm{~G} \nonumber \\
V^r & = & -1.62 \times 10^{11} ~\alpha^{4/5} ~\dot{m}^{1/5} ~x^{-1} \mathrm{~cm~s^{-1}} \nonumber \\
V_A & = & 1.36 \times 10^{13} ~\dot{m}^{1/2} ~x^{-7/4}
\mathrm{~cm~s^{-1}} \label{eqn:mdaf}
\end{eqnarray}
where $x \equiv r / 6 r_G$ at the disk midplane.

Figures \ref{fig:disk_properties} and \ref{fig:disk_properties_2} above include the
predictions for the MDAF transition region (equations
\ref{eqn:mdaf}) over the appropriate radial range, $R_1 < r < R_0$.
We find that the radiatively-cooled numerical simulation T+C fits
the MDAF transition solution remarkably well, except in certain
circumstances.
%, which we discuss below.

First of all, Figure \ref{fig:disk_properties} shows that the density, pressure, and 
$\alpha(r)$ parameter are fit not only
qualitatively but also quantitatively by the analytic transitional
flow model, at least out to $R_0 \approx 150 \, r_G$.  The temperature
structure, however, is a factor of $3$ cooler than the analytic model
and is not as constant with radius.  This is due to the analytic model's
assumption that generic Compton cooling dominates (i.e., $y \approx 1$) when, 
in fact, it is specifically Comptonized {\em synchrotron} cooling that is 
important in the simulations.  The latter can have a slightly different 
value and depends additionally on the magnetic field strength, resulting in 
a slightly lower temperature that may not be constant with radius.
This discrepancy also affects the disk scale 
height, which scales as $\sim \sqrt{T}$.

In Figure \ref{fig:disk_properties_2}, again, most properties
are fit well by the analytic model except for two:  $B^r$
and $rB^{\phi}$.  In the numerical simulation, while the
general magnitude of the magnetic field (seen in $V_A/c$)
fits fairly well (as does even the axial component
$B^{Z}$), the distribution of the rest of the magnetic
field into $r$ and $\phi$ components appears reversed from
the analytic predictions.  That is, the predicted flux
conservation does not take place.  There are different possible
reasons for this:
\begin{itemize}
\item{Any radial shear that could create $B^r$ from $rB^{\phi}$ is suppressed by the 2-D, axisymmetric nature of the
simulations.  3-D simulations may show the expected distribution of $B^r$ and
$B^{\phi}$ in the transition region.}
\item{The natural state of magnetized accretion flow, even with
cooling, may be like that of the RIAF models (e.g., T and I): always
dominated by toroidal magnetic field. In this case, the predicted
MDAF would not arise even with catastrophic cooling.}
\end{itemize}
Therefore, it will be important to compare such 2-D simulations
with similar 3-D ones to see how the inner transitional
and predicted MDAF flows evolve when a third free dimension is
added.  Indeed, both toroidally-dominated and radially-dominated
flows may be possible in nature in this region, with a
state transition between the two occurring from time-to-time.

\section{Conclusions}
\label{sec:conclusions}

Using the Cosmos++ code, we have performed two-dimensional general
relativistic MHD simulations of MRI-unstable accretion flows around
black holes, with the potential of bremsstrahlung, synchrotron, and
Compton cooling of the high-temperature inflow.  In the process of implementing the radiative cooling processes in our code, we made the following observations:
\begin{itemize}
\item{The cooling function in \citet{esi96}, while valid in the range
$10^{8} \, \mathrm{K} < T < 10^{11} \, \mathrm{K}$, needs special
attention and care in order to be valid outside that range and not
lead to heating or freezing runaways in the simulations.}
\item{If radiative cooling is to be added to MRI simulations, then
the energy equation also must properly handle the ``viscous''
heating caused by reconnection and dissipation inherent in the MRI
turbulence. In the present era of moderate-resolution MRI
simulations (where dissipation is caused by numerical effects), this can be handled in one of two ways: perform total-energy-conserving simulations and compute the internal energy by subtracting the
kinetic and magnetic energies from the total; or use an artificial resistivity term to resolve current sheets and allow energy lost through numerical reconnection to be recaptured as heat. Although the artificial resistivity technique has been used with good success in many Newtonian applications \citep[e.g.][]{nit01,sto01,fra05}, it has only recently been tested in a relativistic MHD code \citep{kom07}.}
\item{Indeed, {\em evolving
internal, rather than total, energy without an additional procedure for recapturing lost heat produces an unphysical numerical
cooling} which can rival, or exceed, true radiative cooling. Furthermore, even when the magnitude of cooling is comparable, the resulting disk structure is quite different for the internal-energy-only model.}
\end{itemize}

If our assumption of $T_e \approx T_i$ is valid, then we obtain the following results pertaining to the astrophysics of radiatively cooled, magnetized accretion flows:
\begin{itemize}
\item{Model T+C confirms the ``transitional flow'' solution, which is proposed to connect an outer ADAF-like flow with an inner magnetically-dominated flow, as a viable MHD accretion inflow state.}
\item{The accretion rate and magnetic viscosity parameter
($\dot{M}/\dot{M}_{Edd} \approx 5 \times 10^{-6}$, $\alpha \approx 0.003$)
that result from our choices of inputs are in a range that produces a large transitional flow region, without leading to a completely magnetically dominated state (i.e. $\alpha$ and $\beta^{-1}$ never exceed unity before the flow enters inside
the last stable orbit).}
\item{Comparison of the numerically-computed transitional flow and our
prior analytic models of this region show remarkable qualitative and
quantitative agreement.  Exceptions are limited to the temperature
structure of the analytic model (which used a cooling model much
simpler than the numerical functions herein) and the radial vs. azimuthal magnetic structure (which likely was affected by the
limitations of 2-dimensional axisymmetric MHD). }
\end{itemize}

Further investigations into the development of a true MDAF solution will require additional
2-dimensional simulations to investigate inflows with smaller
transitional and larger predicted true MDAF regions (i.e., with
greater $\alpha$ and $\dot{M}$), and new 3-dimensional simulations
to study the effects of cooling on the ratio of the radial to
toroidal magnetic field components.  The answers to these questions
will determine whether or not radiative cooling ultimately can
trigger the formation of large black hole coronae (MDAFs). These simulations may also help confirm that MDAFs can form moderate-speed jets as proposed in our introduction.

One final point should be noted.  These are some of the first MRI
simulations in which radiative cooling has an important dynamical
effect on the accretion inflow.  It was difficult, therefore, to
predict what the resulting accretion rate and $\alpha$ parameter
would be. The resulting values here ($\alpha = 0.003$, $\dot{M} = 5
\times 10^{-6}$), did not turn out to be appropriate for a source in
the upper right-hand portion of the FBG diagram, as we had originally intended. In fact, they
probably are more appropriate for a source in the {\em lower}
right-hand portion with a rather low accretion rate. The applicability 
of these simulations to such a source, or really any source,
will depend primarily on the validity of the assumption that $T_e = T_i$, as
made in our cooling model.

\begin{acknowledgements}
We thank Sera Markoff and Masa Nakamura for their discussions and careful reading 
of this manuscript. We would like to recognize Joseph Niehaus for his contributions to
testing the Cosmos++ code. PCF gratefully acknowledges the support of a Faculty R\&D
grant from the College of Charleston and a REAP grant from the South
Carolina Space Grant Consortium.  Part of the research described in
this paper was carried out at the Jet Propulsion Laboratory,
California Institute of Technology, under contract to the National
Aeronautics and Space Administration. Part of this research also was
performed when the authors attended extended workshops at the UCLA
Institute for Pure and Applied Mathematics and the UCSB Kavli Institute
for Theoretical Physics.  DLM is grateful to JPL/Caltech for financial
support, and to UCLA for their hospitality during his sabbatical.
This work also was supported by JPL subcontract 1304153.
Computing resources were provided by the JPL Supercomputing Facility and the Texas Advanced Computing Center (TACC).
\end{acknowledgements}

\clearpage
\bibliographystyle{apj}
\bibliography{myrefs}

%%%%%%%%%%%%%%%%%%%%%% FIGURES %%%%%%%%%%%%%%%%%

%%%%%%%%%%%%%%%%%%%% TABLES %%%%%%%%%%%%%%%%%%%%%
%\clearpage

\end{document}